\documentclass[prl,twocolumn,aps,showpacs,amsmath,amssymb]{revtex4-1}

\usepackage{graphicx}
\usepackage{epstopdf}
\usepackage{dcolumn}
\usepackage{bm}
\usepackage{amsmath}
\usepackage{graphicx} 
\usepackage{dcolumn}
\usepackage{bm}
\usepackage{amsmath}
\usepackage{epsfig}
\usepackage{float}
\usepackage{epstopdf}
\usepackage{hyperref}
\hypersetup{breaklinks=true,colorlinks=true,urlcolor=black} 
 
\begin{document}

\title{Discovery of Light-induced Metastable Martensitic Anomaly Controlled by Single-Cycle Terahertz Pulses}

\author
{X. Yang$^{1\dagger}$, B. Song$^{1\dagger}$, C. Vaswani$^{1}$, L. Luo$^{1}$, C. Sundahl$^{2}$, M.~Mootz$^{3}$, J-H. Kang$^{2}$, 
Y. Yao$^{1}$, K-M Ho$^{1}$, I.~E.~Perakis$^{3}$, C. B. Eom$^{2}$ and J. Wang$^{1\ast}$}

\affiliation{$^1$Department of Physics and Astronomy and Ames Laboratory-U.S. DOE, Iowa State University, Ames, Iowa 50011, USA. 
	\\$^2$Department of Materials Science and Engineering, University of Wisconsin-Madison, Madison, WI 53706, USA.
	\\$^3$Department of Physics, University of Alabama at Birmingham, Birmingham, AL 35294-1170, USA.}

\date{\today}

\begin{abstract}
We report on an ultrafast photoinduced phase transition with a strikingly long-lived Martensitic anomaly driven by above-threshold single-cycle terahertz (THz) pulses in 
Nb$_3$Sn. 
A non-thermal, THz-induced depletion of low frequency conductivity indicates  
increased gap splitting of high energy $\Gamma_{12}$ bands by removal of their degeneracies which enhances the Martensitic phase. In contrast, optical pumping leads to a $\Gamma_{12}$ gap melting.   
Such light-induced non-equilibrium Martensitic instability persists up to a critical temperature $\sim$100 K, i.e., more than twice the equilibrium temperature, and can be stabilized beyond technologically-relevant, nanosecond timescales. 
Together with first-principle simulations, we identify a compelling THz tuning of structural fluctuations via E$_u$ phonons to achieve a non-equilibrium ordering at high temperatures far exceeding those for equilibrium states.  
\end{abstract}

\maketitle

\begin{figure}[tbp]
	\includegraphics[scale=0.45]{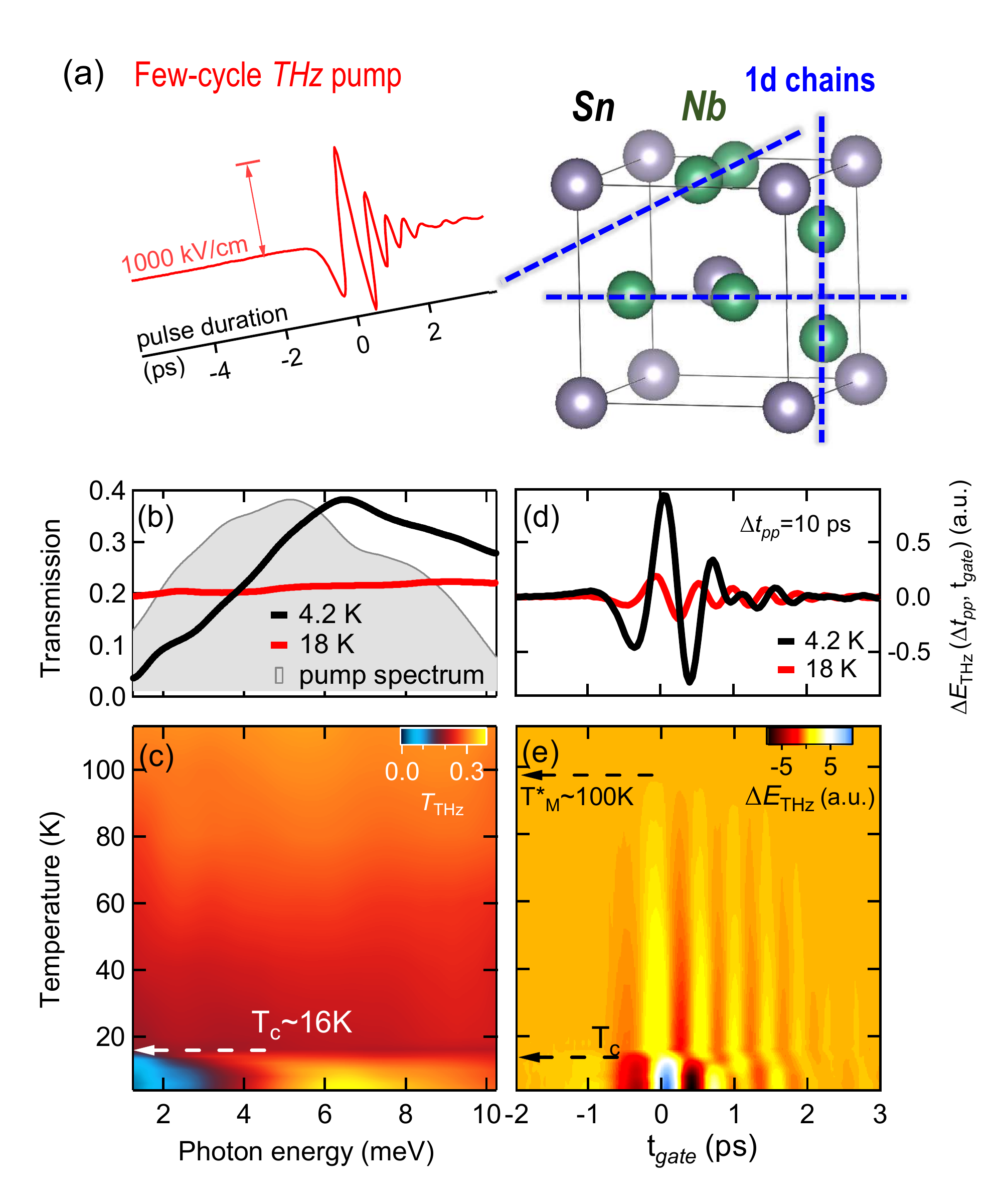}
	\caption{(a) Experimental schematics of driven Martensitic phase with intense THz pulses (red line).  (b) Probe transmission T$_{\mathrm{THz}}(\omega)$ through sample 
		shown together with the pump spectra (gray shade). (c) 2D plot of $t_{\omega, T}$ from 4 K to 110 K. (d), (e) THz Pump-induced transmitted field change $\Delta E_{THz}$ at temperatures same as (b), (c). Dash line marks the SC transition at T$_c=16$ K and THz light-induced critical temperature $T^{\star}_{M}$=100 K. 
	}
	\label{Fig1}
\end{figure}

An emerging paradigm for condensed matter physics is actively explored in light--induced correlation phenomena and phase transitions such as superconductivity \cite{Cavalleri2011Science, YAN19B} and density wave collective orders \cite{more4}. It remains a challenge to develop efficient, non-thermal tuning knobs at THz clock rates and stabilize transient photoinduced phases at many nanosecond timescales.
In contrast to high energy optical excitation, the advent of intense single- and few-cycle THz pulses with peak fields of more than 1000 kV/cm (red line, Fig. 1a) represents a unique opportunity for phase switching and stabilization by applying a lightwave dynamic symmetry breaking principle \cite{SHG} with minimal heating of electronic states \cite{YAN18, Cavalleri2011Nature}. 
A compelling example is the possibility to achieve light--induced superconductivity via multi-THz nonlinear structural pumping in cuprates, which persists far above the equilibrium critical temperatures although it lasts only for few picoseconds \cite{Cavalleri2011Science}. 
Unlike in these intensely-debated, complex materials, A$_3$B compounds, such as Nb$_3$Sn with A15 crystal structure, represent simpler and well-understood model correlated materials \cite{YAN19A, YAN18}, which are well suited for seeking examples of non-equilibrium phase transition by using phonon pumping. Such THz lattice driving is still scarce, despite of recent progress \cite{Cavalleri2011Science, MKoz19,Sie19, THz1, THz2, THz3,Nelson19}.
Here we use Nb$_3$Sn to address two outstanding general issues: (i) can intense THz light pump fields create {\em long-lived} Martensitic orders {\em far above} equilibrium critical temperatures? (ii) what are the salient features of the non-thermal tuning of structural fluctuations that give rise to such controllable non-equilibrium order? 

A Martensitic normal state transition in Nb$_3$Sn can be understood 
in terms of electronic and structural instabilities, which associate with 
optical phonon condensation. As illustrated in Fig. 1a, ``dimerization'' of Nb atoms emerges along three one-dimensional chains (blue dash lines) at $T_M$=48~K above the superconducting transition at $T_c$~\cite{Shirane, Kataoka1983, Bilbro1976,Sadigh1998}. Such structural (cubic-tetragonal) and phonon softening anomalies can originate from a Van Hove singularity (VHS)-like, electronic density-of-states (DOS) peaked at $\sim$E$_F$ and from strong electron-phonon interaction. 
These give rise to a Jahn-Teller effect due to two fold-degenerate $\Gamma_{12}$ sub-bands crossing the Fermi level, with DOS that determines $T_{M}$ of the Martensitic phase transition.    
Therefore, the $\Gamma_{12}$ phonon pumping by an intense, few-cycle THz-pulse $\sim$1000 kV/cm (red trace, Fig. 1a), without significantly heating of other degrees of freedom, provides a compelling avenue to induce a non-equilibrium Martensitic phase at temperatures far exceeding the equilibrium $T_{M}$, by lifting the electronic degeneracy and increasing DOS of the $\Gamma_{12}$ bands. 
Although quantum quench of superconducting states has been actively explored in Nb$_3$Sn \cite{YAN18}, the THz-driven Martensitic normal states have never been explored which is the focus of this work.

In this letter, we present a light-induced metastable, Martensitic phase out-of-equilibrium in Nb$_3$Sn obtained by single-cycle THz pumping. 
The photoinduced non-equilibrium Martensitic phase displays the non-thermal electrodynamics that persist up to a critical $T^{\star}_{M}\sim$2$T_M$, i.e., doubling of the equilibrium value, for longer than 1 ns. 
Our theoretical modeling underpins a $\Gamma_{12}$ phonon-tuning mechanism of the Martensitic instability and explains, particularly, the doubling of T$_M$ and non-thermal conductivity depletion. 

The sample measured in the experiment is a 20nm Nb$_3$Sn film grown on (100) oriented sapphire single crystalline substrates by pulsed laser deposition. 
Single cycle THz pump pulses were generated by a tilted-pulse-front phase matching through 1.3\% MgO doped LiNbO3 crystal. Peak E field is as large as 1000 kV/cm (Fig. 1a) and spectrum (gray shade, Fig. 1b) covers $\sim$1-10 meV.  
Complex transmission $\tilde{t}(\omega)$ is obtained by 
Fourier spectra of transmitted THz probe field oscillation in time domain. 
Frequency dependent optical conductivity $\sigma_1(\omega)$ and $\sigma_2(\omega)$ extracted from $\tilde{t}(\omega)$ measures dissipative and inductive response respectively \cite{last1,last2,last3,last4,last5}. 

\begin{figure}[tbp]
	\includegraphics[scale=0.38]{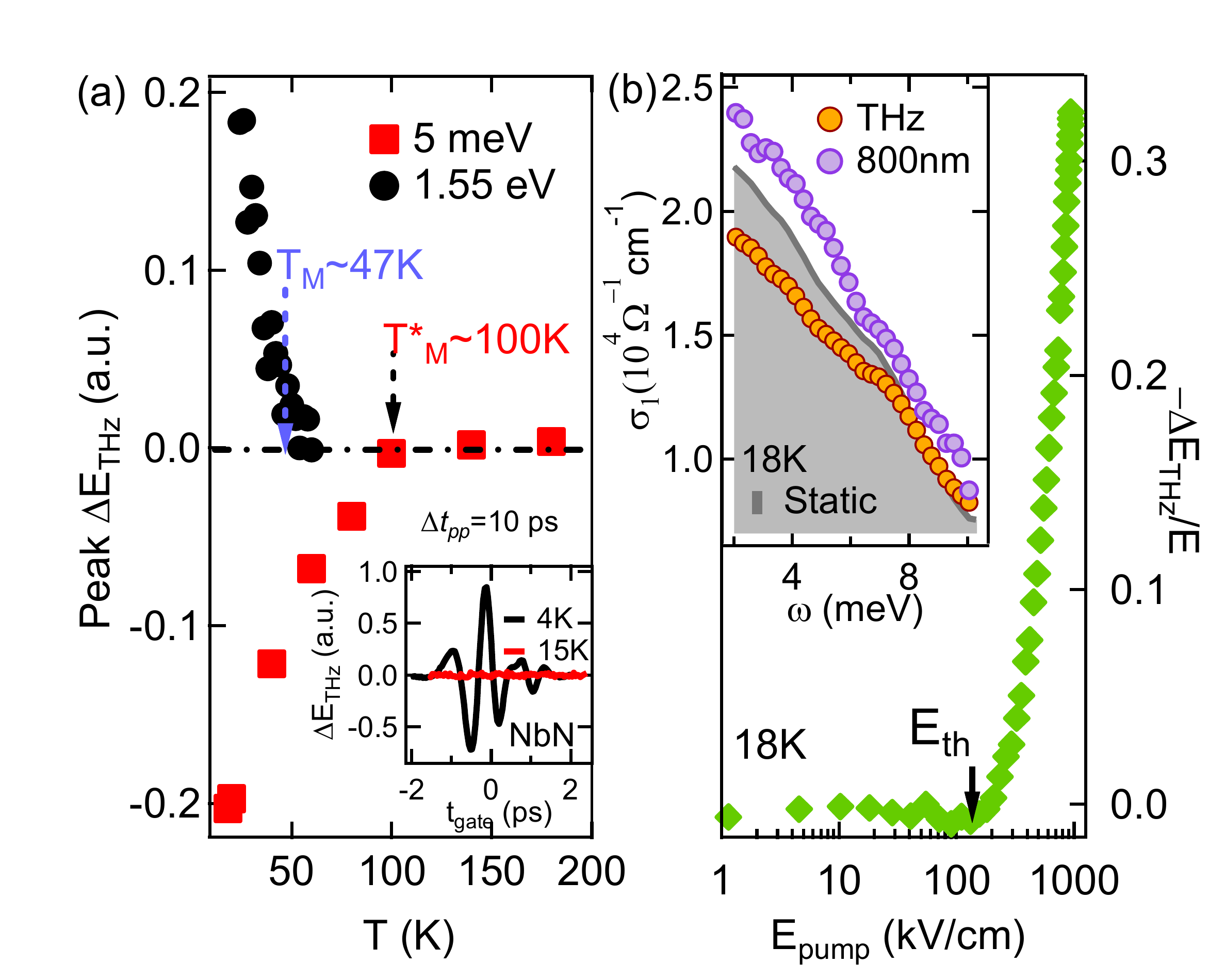}
	\caption{
		(Color online). (a) Temperature dependence of pump induce change $\Delta \mathrm{E}_{\mathrm{THz}}$ under THz and 1.55eV optical photo-excitation. Martensitic transition temperature T$_M$ and $T^{\star}_{M}$ for vanishing $\Delta \mathrm{E}_{\mathrm{THz}}$ under THz pump is marked by purple and black arrows, respectively. Inset: No pump induced change is observed in a NbN superconductor above $T_c=14K$. (b) Pump field dependence of $\Delta E_{\mathrm{THz}}$ at 18K shows a threshold $E_{th}$ at 130kV/cm. Inset: Non-equilibrium conductivity $\sigma_1(\omega)$ under THz and 1.55 eV pump compared to thermal equilibrium state at 18K. 
	}
	\label{Fig2}
\end{figure}

Fig.1b presents the static THz transmission of Nb$_3$Sn $T(\omega)=|\tilde{t}(\omega)|$ at 4.2K and 18K. 
quasi-particle (QP) excitation gap $2\Delta_{SC}$ gives rise to the SC state line shape (black line), while the normal state spectrum (red line) is largely featureless, tilting slightly up towards high frequency. 
A 2D false-color plot of THz transmission spectra $T(\omega)$ at various temperatures (Fig.1c) shows distinctly different shapes below and above critical temperature T$_c\sim$16K. 
There the transmission peak diminishes and redshifts with increasing temperature, and completely vanishes when approaching T$_c$. 
Our focus next is THz pumping of Martensitic normal states above T$_c$. Fig. 1c shows that equilibrium transmission spectra above T$_c$ show very little changes in the measured frequency range, i.e., the static THz conductivity is not very sensitive to the Martensitic normal state order.

In strong contrast, non-equilibrium signals after THz pump centered at $\sim$ 5 meV show a clear temperature-dependence in the normal state.  
Fig.1d shows typical pump-induced changes of transmitted field $\Delta \mathrm{E}_{\mathrm{THz}}$ in time domain at $\Delta t_{pp}=10ps$. 
The normal state, 18K trace shows a clearly phase shift and amplitude reduction in comparison with the 4.2K trace.       
Fig.1e presents a 2D false plot of $\Delta \mathrm{E}_{\mathrm{THz}}$ up to $\sim$110K.
Again, normal state is well separated from SC state
across T$_c$. 
Most intriguingly, clear $\Delta \mathrm{E}_{\mathrm{THz}}$ signals are detected in the normal state and persist up to 100 K, i.e., $T^{\star}_{M}\sim$2$T_M$, indicative of a photo-induced non-equilibrium order far above $T_M$.

To further underpin the THz pump-induced $T^{\star}_{M}$ phase, Fig.2a shows pump-induced differential transmission $\Delta \mathrm{E}_{\mathrm{THz}}$ under conventional optical pump, 1.55eV. Here the optically-induced THz signals (black circles) vanish at the equilibrium transition at T$_M$=48K. 
These results clearly establish ultrafast optically-induced $\Delta \mathrm{E}_{\mathrm{THz}}$ as an effective probe for the equilibrium Martensitic order. 
In contrast, the 5meV photo-excitation clearly establishes the non-zero $\Delta \mathrm{E}_{\mathrm{THz}}$ signals up to $T^{\star}_{M}$, indicative of ``order-parameter-like" response for the non-equilibrium Martensitic phase. 

Further experimental evidence associating the THz-driven phase transition is presented in Fig. 2b, which plots THz pump field dependence of $\Delta \mathrm{E}_{\mathrm{THz}}$ signals at a fixed time $\Delta t_{pp}=$10ps. It is clearly visible that the signal is negligibly small at THz field strengths less than E$_{\mathrm{th}}\sim$130kV/cm but increases significantly above it. Such distinct threshold behavior of the THz-driven dynamics is not limited by our noise floor, which is a hallmark of the non-equilibrium phase transition to an induced Martensitic phase. For comparison, no THz pump-induced change is observed in a NbN superconductor in the normal state without the Martensitic order, e.g., the 15K trace in (red line, inset, Fig. 2a) for NbN vs the 18K trace (red line, Fig. 1d) for Nb$_3$Sn.   

The non-equilibrium response function $\sigma_1(\omega)$ in Fig. 2(b) (inset) reveals different behaviors for 1.55eV (optical, purple) and 5meV (THz, orange) pumping, which distinguish thermal vs non-thermal electrodynamics. 
After high photon energy, 1.55eV pump excitation, the low frequency conductivity gains an additional spectral weight over its equilibrium (no pump) values (gray shade), i.e., $\Delta \sigma_1(\omega)>$0. This can be understood as melting of the high energy electronic gaps that develop at the T$_M$ transition from $\Gamma_{12}$ phonon condensation (dimerization). The latter leads to spectral weight transfer to the Fermi surface by suppressing the Martensitic phase via hot phonons and electrons excited by the high energy photons.  
However, the 5meV pump photon energy is far below the gap and thus cannot quench it. Instead, it reverses the spectral weight transfer to high energy by reducing $\sigma_1(\omega)$. 
This is consistent with an elevated transition temperature $T^{\star}_{M}$ and, thereby, enhanced non-equilibrium Martensitic order by acquiring an extra spectral weight from the Fermi surface, as quantitatively substantiated latter. 

\begin{figure}[tbp]
	\includegraphics[scale=0.37]{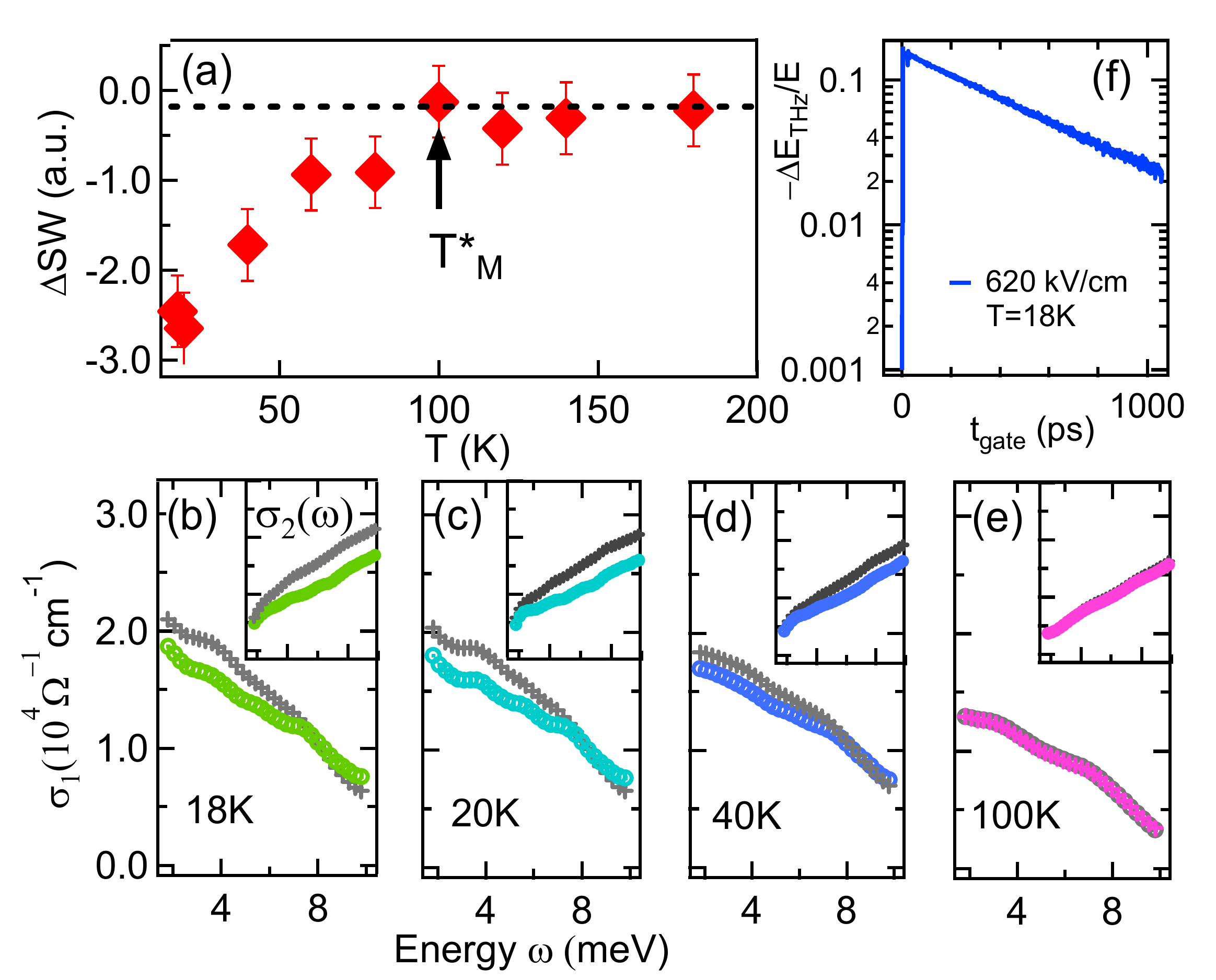}
	\caption{
		(Color online). (a) Spectral weight change $\Delta SW$ integrated from 2-10meV under highest driving field of THz excitation. (b)-(e): Real part of conductivity $\sigma_1(\omega)$ compared to thermal equilibrium at various temperatures with $\sigma_2(\omega)$ shown in inset. (f) $\Delta E_{THz}$/E temporal dynamics at $E_{THz}$=620kV/cm and 18K.  
	}
	\label{Fig2}
\end{figure}

Fig. 3 presents extensive conductivity spectra measurement and spectral weight (SW) analysis in the normal state to investigate Martensitic dynamics under intense THz radiation.
The integrated SW change $\Delta \mathrm{SW}$ (1-10 meV) induced by the THz pumping is shown in Fig. 3a at various temperatures in the normal state, together with complex conductivity spectra, $\sigma_1(\omega)$ and $\sigma_2(\omega)$, shown in Figs. 3b-3e.
The most salient feature is 
the depleted SW, i.e., $\Delta SW < 0$, that emerges elusively below $T^{\star}_{M}$. 
Such SW removal in $\sigma_1(\omega)$ corresponds to a reduction in $\sigma_2(\omega)$ (inset) as compared to thermal-equilibrium states (gray crosses, Figs. 3b-3e), which is correlated by Kramers-Kronig transformation. 
Since the total integrated SW is conserved, the missing spectral component is expected to transfer to high energy electronic states beyond the measurement energy window, which gives rise to a transient increase of $T_M$ to $T^{\star}_{M}$. 
Such experimental evidence indicates strong correlation between THz-controlled SW transfer and Martensitic order. 
Furthermore, the $T^{\star}_{M}$ phase appears to be metastable, as witnessed by the long $\sim$1\,~ns relaxation time, e.g., as shown in 620kV/cm trace (black line) in Fig. 3f.  

\begin{figure*}[!tbp]
	\includegraphics[scale=0.6]{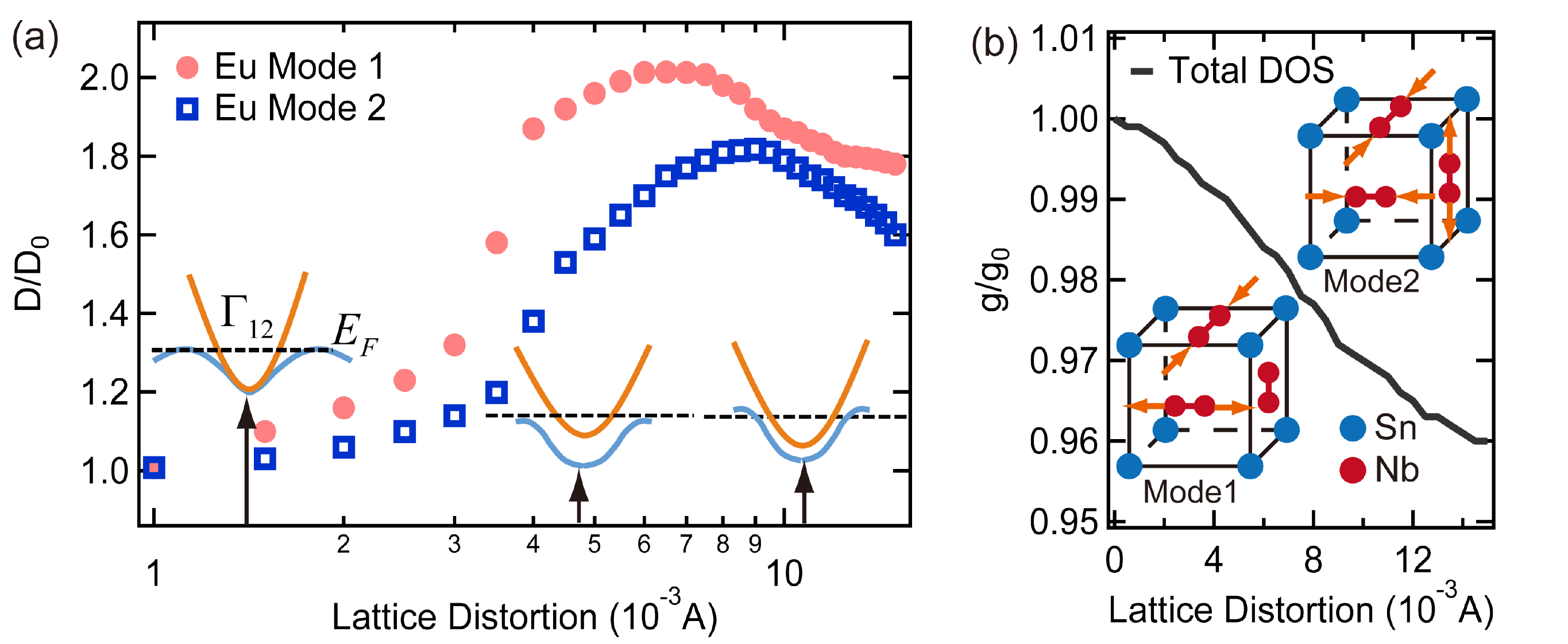}
	\caption{\label{fig4} (Color online). (a) DOS ($D$) of two equivalent $\Gamma_{12}$ electronic bands and (b) total DOS ($g$) at Fermi level under finite lattice distortions (compared to its equilibrium $D_0$, $g_0$). Inset in (a): the schematic of $\Gamma_{12}$ bands under lattice distortion. Inset in (b): Two E$_u$ phonon modes are manifested as lattice vibration of Nb atoms in chains along three different directions.
	}
\end{figure*}

The effects of THz field interaction with the Martensitic phase in Nb$_3$Sn can be understood by 
considering degenerate $\Gamma_{12}$ electronic bands crossing the Fermi level (inset, Fig. 4a) and strong electron-phonon coupling \cite{Gorkov2013, Bhatt1976, Bilbro1976, Bhatt1977, Kataoka1983}. Two degenerate E$_u$ phonon modes involved are shown in Fig. 4b (inset).
It has been proposed that the transition temperature $T_{M}$ is proportional to the DOS of ${\Gamma}_{12}$ electronic bands at the Fermi level \cite{Kataoka1983}.
Our physical picture and simulations below demonstrate that $\Gamma_{12}$ lattice vibration driven by low energy photo-excitation close to E$_u$ phonon resonances is able to lift the degeneracy and modify the DOS of the $\Gamma_{12}$ band, as illustrated in Fig. 4a (inset). 
Active modulations of the Martensitic transition temperature, correlation gap and electronic order are achieved by intense THz radiation, which is absent for high energy, optical pumping with photon energy far above the E$_u$ phonon resonances.  
The $\Gamma_{12}$ lattice vibration most relevant to the Martensitic phase can be understood as dimerization of the Nb atom chain along different axes (E$_u$ symmetry, inset, Fig. 4b) similar to a charge density wave but confined to a single unit cell with wave vector $q=0$.

To put the above-mentioned physical picture on a sound footing, we lay out first-principle simulations of the phonon-assisted tuning of the Martensitic phase (see Supplementary for details). 
The $\Gamma_{12}$  phonon contains two degenerate modes at $\Gamma$ point in B.Z., which are (deduced from symmetry)
\begin{equation}
\begin{split}
Q_1&=\frac{A_1}{2}(-u_{2x}+u_{1x}+u_{4y}-u_{3y}) \\
Q_2&=\frac{A_2}{2\sqrt{3}}(2u_{6z}-2u_{5z}-u_{2x}+u_{1x}-u_{4y}+u_{3y})
\end{split}
\label{eq1}
\end{equation}
$u_{i\sigma}$ stand for the displacement of $i^{th}$ Nb atoms in the $\sigma$ Cartesian component. The $x$-coordinates in Fig.~4 correspond to $A_1$  and $A_2$  in the above equations. 

The total free energy can be constructed based on the two E$_u$ modes
\begin{equation}
\begin{split}
F=\frac{1}{2}Vc_0u^2+\frac{1}{2}{\omega}^2(Q_1^2+Q_2^2)+{\zeta}\sqrt{Vc_0}{\cdot}{\omega}uQ_1\\
+Vn{\mu}-2k_{B}T(S_{b1}+S_{b2}),
\end{split}
\label{eq2}
\end{equation}
where $V$ is unit cell volume, $c_0$ is force constant, $\mu$ is chemical potential, and $u$ is defined by
\begin{equation}
u=(2e_{zz}-e_{xx}-e_{yy})/\sqrt{6},
\label{eq3}
\end{equation}
where $e_{xx}$, $e_{yy}$, $e_{zz}$ are the diagonal components of strain tensors. $S_{b1}$ and $S_{b2}$ are the entropy due to the two $\Gamma_{12}$ electronic bands, which are expressed as
\begin{equation}
S_{iB}=\sum_{k}ln[1+exp[-({\varepsilon}_k^i-{\mu})]/k_{B}T]
\label{eq3}
\end{equation}
The zero order $H_0$ of band energies ${\varepsilon}_k$ can be simplified as parabolic, which are plausible approximations for dispersion near the $\Gamma$ point in B.Z. It will perturbed by the following Hamiltonian $H_e=H_0+H'$

\begin{equation}
\begin{split}
H'&=[\frac{{\eta}_i{\omega}}{\sqrt{nV}}Q_2+\frac{{\hbar}^2}{2m} \frac{1} {\sqrt{2}}(k_{x}^2-k_{y}^2)](c_1^{\dagger}c_2+c_2^{\dagger}c_1)\\
&+[{\eta}_0\sqrt{\frac{c_0}{n}}u+\frac{{\eta}_i{\omega}}{\sqrt{nV}}Q_1+\frac{{\hbar}^2}{2m}\frac{1}{\sqrt{6}}(3k_z^2-k^2)](c_2^{\dagger}c_2-c_1^{\dagger}c_1)
\end{split}
\label{eq4}
\end{equation}
In above, we have ignored the $k$ index for $c_1$ $c_2$, etc. Note that ${\eta}_0$, ${\eta}_i$ come into $F$ through ${\varepsilon}^i_k$.

In order to account for the Martensitic transition, the model has included several degrees of freedom: the elastic distortion (tensor) $u$, optical phonon modes $Q_1$ $Q_2$, electron entropy, and $e$-phonon coupling. Minimize $F$ with respect to $u$, $Q_1$, $Q_2$ and $\mu$, yielding the equilibrium lattice displacement, which relies on numerical solutions. It shows that the critical $T_M$ is approximately proportional with a parameter ${\alpha}$ in a broad regime, i.e., $k_{B}T_{M}/{\varepsilon}_F~{\propto}~{\alpha}$.
\begin{equation}
\begin{split}
{\alpha}&=2a_0^3D({\varepsilon}_F)G_0^2,\\
G_0&=|{\eta}_0|\sqrt{1+\frac{({\eta}_i-{\zeta}{\eta}_0)^2}{(1-{\zeta}^2){\eta}_0^2}}
\end{split}
\label{eq3}
\end{equation}
$D({\varepsilon}_F)$ is ${\Gamma}_{12}$ density of states at Fermi level, and $a_0$ is the lattice parameter of undistorted unit cell (cubic). Notice that $a_0$ and $G_0$ could largely be taken as constant, then it yields, $T_M~{\propto}~{\alpha}~{\propto}~D({\varepsilon}_F)$. 

Next, we explicitly evaluate ${\Gamma}_{12}$-DOS change with presence of ${\Gamma}_{12}$ phonon modes.
The electronic structure under $\Gamma_{12}$ lattice distortion is examined by density functional theory (DFT) \cite{Kessel1978} with PAW methods \cite{Blochl} and Perdew-Burke-Ernzerhof (PBE) exchange-correlation functional \cite{Matteiss1982} using the ``Frozen phonon" approximation. 
Our results are presented in Fig.4a, which shows the DOS of $\Gamma_{12}$ electronic bands at the Fermi level (FL) vs phonon amplitudes of the characterized by Nb atom displacement from its equilibrium position. Note that Fig.4a only accounts for the DOS due to the $\Gamma_{12}$ bands, which is different from total DOS (Fig. 4b).
Strikingly, DOS of $\Gamma_{12}$ bands undergoes a sharp increase above threshold at 0.002$\AA$ and reaches a maximum two-time enhancement at displacement as small as 0.006$\AA$, followed by a slight drop under further distortion. The sudden increase is due the band (blue in Fig. 4a) that touches and crosses the FL at a critical amplitude, leading to an approximate doubling of the $\Gamma_{12}$-DOS. 
Thus, simulations well address the experimental observation of two-time enhancement in transition temperature $T^* \simeq 2T_M$ and also implies a threshold E field ($\sim$130 kV/cm) for enhanced $T_M$, as observed in Fig. 2a.
On the other hand, in Fig. 4b, the decrease in the total DOS, i.e., $\sim$4\% in the whole range of phonon distortions, is in excellent agreement with the decreased SW in THz driven states below $T^{\star}_{M}$ observed in Figs. 3b-3e. Furthermore, our DFT calculation shows E$_u$ phonon energy $\sim$ 12 meV, which is an overestimate since it cannot fully capture electron-phonon interaction near the Martensitic transition \cite{Tutuncu2006}. Nevertheless, the E$_u$ phonon energy is clearly close the THz pump pulse up to $\sim$10 meV, which can excite the ${\Gamma}_{12}$ resonance non-thermally.  


In summary, we demonstrate a light-enhanced Martensitic phase driven by intense, single-cycle THz fields, manifested as the doubling of transition temperature and removal of spectral weights in the vicinity of the Fermi level. 
First-principle calculations reveal an effective non-thermal modulation of degenerate $\Gamma_{12}$ electronic bands that determine the Martensitic phase and consistently explain all the key experimental features.  
The light-induced phonon tuning can be extended to topological matter~\cite{more1}, 2D materials~\cite{2D}, magnetism~\cite{more2,spin2} and unconventional superconductors~\cite{more3, Patz2018}.
Our work also provides compelling implications for quantum computation applications since doped Nb$_3$Sn is still the material of choice to replace Al-based transmon qubits and support high current/magnetic field applications, despite of much improved $T_c$ in unconventional superconductors. 

\begin{acknowledgments} 
	This work was supported by National Science Foundation 1905981 (THz spectroscopy). 
	B.S. and L.L. (DFT calculation and model building) were supported by the U.S. Department of Energy, Office of Basic Energy Science, Division of Materials Sciences and Engineering (Contract No. DE-AC02-07CH11358).
	Work at the University of Wisconsin was supported by the Department of Energy Office of Basic Energy Sciences under award number DE-FG02-06ER46327 (structural and electrical characterizations) and Department of Energy Grant no.  DE-SC100387-020 (sample growth).
	Data analysis work at the University of Alabama, Birmingham was supported by the US Department of Energy under contract \# DE-SC0019137 (M.M and I.E.P).
	The THz Instrument was supported in part by National Science Foundation EECS 1611454.
	
	\noindent$^{\ast}$Corresponding author: jwang@ameslab.gov.\\
	\normalsize{$\dagger$ Equal contribution}
\end{acknowledgments}


\begin{thebibliography}{99}
	
	\bibitem{YAN19B} Yang, X. et al. 
	\textit{Nat. Photon.} \textbf{13}, 707 (2019)

\bibitem{Cavalleri2011Science} Fausti, D. \textit{et al.} \textit{Science} \textbf{331}, 189-192 (2011).

\bibitem{more4} Patz, A. \textit{et al.} 
	\textit{Nat. Commun.} \textbf{5}, 3229 (2014).
	
	\bibitem{SHG} C. Vaswani \textit{et al}., 
	\textit{Phys. Rev. Lett.} \textbf{124}, 207003 (2020).

\bibitem{YAN18} Yang, X. et al. 
	\textit{Nat. Mater.} \textbf{17}, 586-591 (2018)

\bibitem{Cavalleri2011Nature} Dienst, A. \textit{et al.} \textit{Nat. Photonics.} \textbf{5}, 485-488 (2011).

\bibitem{YAN19A} Yang, X. et al. 
	\textit{Phys. Rev. B} \textbf{99}, 094504 (2018)

\bibitem{MKoz19} M. Kozina, M. Fechner, P. Marsik, T. van Driel, J. M. Glownia, C. Bernhard, M. Radovic, D. Zhu, S. Bonetti, U. Staub, M. C. Hoffmann, 
\textit{Nat. Phys.}, \textbf{15}, 387 (2019) 

\bibitem{Nelson19} X. Li, T. Qiu, J. Zhang, E. Baldini, J. Lu, A. M. Rappe, and K. A. Nelson, 
\textit{Science}, \textbf{364}, 1079 (2019)

\bibitem{Sie19} E. Sie et al., 
Nature \textbf{565}, 61 (2019).

\bibitem{THz1} Vaswani, C. et al., 
	\textit{Phys. Rev. X} \textbf{10}, 021013 (2020)

\bibitem{THz2} Liu, Z. et al., 
	\textit{Phys. Rev. Lett} \textbf{124}, 157401 (2020)

\bibitem{THz3} X. Yang et al., 
	\textit{npj Quantum Mater.}, \textbf{5}, 13 (2020). https://doi.org/10.1038/s41535-020-0215-7
	
\bibitem{Shirane} Shirane, G. and Axe, J. D. 
	\textit{Phys. Rev. B} \textbf{4}, 2957 (1971).

\bibitem{Bilbro1976} G. Bilbro, and W. L. McMillan, Phys. Rev. B, \textbf{14}, 1887 (1976)
	 
\bibitem{Kataoka1983} M. Kataoka, Phys. Rev. B, \textbf{28}, 2800 (1983)

\bibitem{Sadigh1998} B. Sadigh, and V. Ozoliņs. Phys. Rev. B, \textbf{57}, 2793 (1998)
	
\bibitem{last1} Luo, L. \textit{et al.} 
\textit{Nat. Commun.} \textbf{5}, 3055 (2014).
	
\bibitem{last2} Luo, L. \textit{et al.} 
\textit{Nat. Commun.} \textbf{8}, 15565 (2017).	

\bibitem{last3} Wang, J. \textit{et al.} 
\textit{Phys. Rev. Lett.,} \textbf{104}, 177401 (2010).	

\bibitem{last4} Luo, L. \textit{et al.} 
\textit{Phys. Rev. Lett.,} \textbf{114}, 107402 (2015).	

\bibitem{last5} Luo, L. \textit{et al.} 
\textit{Phys. Rev. Materials,} \textbf{3}, 026003 (2019).

	
\bibitem{Gorkov2013} L. P. Gor'kov, Inst. of Theoretical Physics, Moscow, \textbf{65}, 1658 (1973)

\bibitem{Bhatt1976}  R. N. Bhatt, and W. L. McMillan. Phys. Rev. B, \textbf{14}, 1007 (1976)

\bibitem{Bhatt1977} R. N. Bhatt, Phys. Rev. B, \textbf{16}, 1915 (1977)

\bibitem{Kessel1978} A. T. Van Kessel, H. W. Myron, and F. M. Mueller. Phys. Rev. Lett. \textbf{41}, 3. 181 (1978).

\bibitem{Blochl} P. E. Blochl, Phys. Rev. B \textbf{50}, 17953 (1994)

\bibitem{Matteiss1982} L.F. Mattheiss, and W. Weber. Phys. Rev. B, \textbf{25}, 2248 (1982)

\bibitem{Tutuncu2006} H. M. Tutuncu, G. P. Srivastava, S. Bagci, and S. Duman.  Phys. Rev. B \textbf{74}, 212506 (2006).

\bibitem{more1} L. Luo et al., 
	\textit{Nat. Commun.} \textbf{10}, 607 (2019).

\bibitem{2D} See, e.g., T. Li, et al. 
	Phys. Rev. Lett. \textbf{108}, 167401 (2012).

\bibitem{more2} T. Li et al., 
	Nature {\bf 496}, 69 (2013) 	
	
\bibitem{spin2} A. Patz, T. Li, X. Liu, J. K. Furdyna, I. E. Perakis, and J. Wang, 
	\emph{Phys. Rev. B }\textbf{91}, 155108 (2015).

\bibitem{more3} X.Yang et al., 
	\textit{Phys. Rev. Lett.} \textbf{121}, 267001 (2018) 
	
\bibitem{Patz2018} Patz, A. et al. \textit{Phys. Rev. B} \textbf{95}, 165122 (2017)

	
\end{thebibliography}
\end{document}